\documentclass[fleqn,usenatbib]{mnras}

\usepackage[T1]{fontenc}

\DeclareRobustCommand{\VAN}[3]{#2}
\let\VANthebibliography\thebibliography
\def\thebibliography{\DeclareRobustCommand{\VAN}[3]{##3}\VANthebibliography}

\usepackage{graphicx}	
\usepackage{amsmath}	
\usepackage{longtable}
\usepackage[flushleft]{threeparttable}
\usepackage{tabularx}
\usepackage{multirow}
\usepackage{graphicx}
\usepackage{amsmath,amssymb}
\usepackage{color}
\usepackage{units}
\usepackage{epstopdf}
\usepackage{hyperref}
\usepackage{multirow}
\usepackage{url}
\usepackage{subfigure}
\usepackage{rotating}
\usepackage{enumitem}\setlist[description]{font=\textendash\enskip\scshape\bfseries}
\usepackage{lineno}
\usepackage[table,xcdraw]{xcolor}

\newcommand{\beq}{\begin{equation}}
\newcommand{\eeq}{\end{equation}}
\newcommand{\bdm}{\begin{displaymath}}
\newcommand{\edm}{\end{displaymath}}

\definecolor{Gray}{gray}{0.9}
\definecolor{orange}{rgb}{0.9,0.5,0}

\newcommand{\ztfrest}{\text{ZTFReST}}

\newcommand{\nmma}{\text{NMMA}}

\usepackage{txfonts}

\title[An Online Framework for Fitting Fast Transient Lightcurves]{An Online Framework for Fitting Fast Transient Lightcurves}

\author[Barna et al.]{Tyler Barna,$^{1}$
Brandon Reed,$^{2}$
Igor Andreoni,$^{3,4,5}$
Michael W. Coughlin,$^{1}$
Tim Dietrich,$^{6,7}$
\newauthor
Steven L. Groom,$^{8}$
Theophile Jegou du Laz,$^{9}$
Peter T.~H.~Pang,$^{10,11}$
Josiah N. Purdum,$^{12}$
\newauthor
and Ben Rusholme$^{8}$
\\ 
$^{1}$ School of Physics and Astronomy, University of Minnesota, Minneapolis, Minnesota 55455, USA\\
$^{2}$ Department of Physics and Astronomy, University of Minnesota -- Duluth, Duluth, Minnesota 55812, USA\\
$^{3}$ Joint Space-Science Institute, University of Maryland, College Park, MD 20742, USA\\
$^{4}$ Astrophysics Science Division, NASA Goddard Space Flight Center, MC 661, Greenbelt, MD 20771, USA\\
$^{5}$ Department of Astronomy, University of Maryland, College Park, MD 20742, USA\\
$^{6}$ Institute of Physics and Astronomy, Theoretical Astrophysics, University Potsdam, Haus 28, Karl-Liebknecht-Str. 24/25, 14476, Potsdam, Germany\\
$^{7}$ Max Planck Institute for Gravitational Physics (Albert Einstein Institute), Am Mühlenberg 1, Potsdam 14476, Germany\\
$^{8}$ IPAC, California Institute of Technology, 1200 E. California Blvd, Pasadena, CA 91125, USA\\
$^{9}$ Division of Physics, Mathematics, and Astronomy, California Institute of Technology, Pasadena, CA 91125, USA\\
$^{10}$ Nikhef, Science Park 105, 1098 XG Amsterdam, The Netherlands\\
$^{11}$ Institute for Gravitational and Subatomic Physics (GRASP), Utrecht University, Princetonplein 1, 3584 CC Utrecht, The Netherlands\\
$^{12}$ Caltech Optical Observatories, California Institute of Technology, Pasadena, CA 91125, USA\\
}

\pubyear{2024}

\begin{document}
\label{firstpage}
\pagerange{\pageref{firstpage}--\pageref{lastpage}}
\maketitle


\begin{abstract}
The identification of extragalactic fast optical transients (eFOTs) as potential multi-messenger sources is one of the main challenges in time-domain astronomy. However, recent developments have allowed for probes of rapidly-evolving transients. With the increasing number of alert streams from optical time-domain surveys, the next paradigm is building technologies to rapidly identify the most interesting transients for follow-up. One effort to make this possible is the fitting of objects to a variety of eFOT lightcurve models such as kilonovae and $\gamma$-ray burst (GRB) afterglows. In this work, we describe a new framework designed to efficiently fit transients to light curve models and flag them for further follow-up. We describe the pipeline's workflow and a handful of performance metrics, including the nominal sampling time for each model. We highlight as examples ZTF20abwysqy, the shortest long gamma ray burst discovered to date, and ZTF21abotose, a core-collapse supernova initially identified as a potential kilonova candidate.
\end{abstract}




\section{Introduction}

The detection of GW170817 \citep{AbEA2017b} and its associated electromagnetic transients AT2017gfo \citep{CoFo2017,SmCh2017,AbEA2017f} and GRB170817A~\citep{GoVe2017,SaFe2017,AbEA2017e} has accelerated the field of multi-messenger astronomy. In particular, this single event has increased our knowledge of the neutron star equation of state \citep{BaJu2017, MaMe2017, CoDi2018b, CoDi2018, CoDi2019b, AnEe2018, MoWe2018,RaPe2018,Lai2019,DiCo2020}, the Hubble constant \citep{CoDi2019,CoAn2020,2017Natur.551...85A,HoNa2018,DiCo2020}, and $r$-process nucleosynthesis \citep{ChBe2017,2017Sci...358.1556C, CoBe2017,PiDa2017,RoFe2017,SmCh2017,WaHa2019,KaKa2019}. However, with the third LIGO-Virgo observing run (O3) ending without a viable optical counterpart to a binary neutron star or neutron star--black hole merger candidate \citep[e.g.,][]{Andreoni2019S190510g, CoAh2019b, Goldstein2019S190426c, Gomez2019, LuPa2019, AnCo2020, Ackley2020, Andreoni2020S190814bv, AnAg2020,GoCu2020,KaAn2020}, improving upon the technology we use to efficiently identify transient candidates for follow-up is necessary as we look to the fourth LIGO-Virgo observing run (O4) and beyond. 

During the down-time between the end of O3 and the start of O4, a new framework for fast transient identification known as ZTF Realtime Search and Triggering (\ztfrest) \citep{AnCo2021} was developed. This framework has already identified at least seven confirmed afterglows and early identification of a number of these sources has made it possible to carry out multi-wavelength follow-up observations.
The development of this framework was motivated by searches for serendipitous kilonovae in optical survey data, which rapidly fade in the optical and therefore require online frameworks to have hope for successful identifications.

Following \ztfrest, the Nuclear-physics and Multi-Messenger Astrophysics (\nmma) framework \citep{Pang2022} was introduced. There are a number of transient classes that possess similar features to kilonova at early times, making the optimization of follow-up observations difficult. To address this, \nmma \ was developed, which allows for Bayesian inference on EM observations to classify objects. In the event that another multi-messenger event occurs, \nmma \ also enables researchers to perform combined analyses of EM and GW signals to constrain the Hubble Constant and neutron star equation of state (EoS).

In this paper, we describe an improvement of the \ztfrest \ automated infrastructure through the inclusion of multi-modal EM lightcurve fitting using \nmma \ rather than the optional kilonova model fitting as described in Section 2 of \cite{AnCo2021}. In addition, we have developed an extension to the pipeline that allows for the lightcurve analyses to be easily accessed by the broader collaboration in a Slack channel as well as access to these analyses via Fritz.
We describe the pipeline in Sec.~\ref{sec:pipeline}.
Example results and an assessment of \nmma \ is shown in Sec.~\ref{sec:results}.
We summarize our conclusions and future outlook in Sec.~\ref{sec:discussion}.

\section{Pipeline Framework}\label{sec:pipeline}
This online framework builds upon \ztfrest\footnote{The \ztfrest \ repository can be found \href{https://github.com/growth-astro/ztfrest}{here}.} and \nmma\footnote{The \nmma \ repository can be found \href{https://github.com/nuclear-multimessenger-astronomy/nmma}{here}} in order to deliver automated fitting of candidates on a regular basis. Additional information regarding the methods underlying \ztfrest \ and \nmma \ can be found in \cite{AnCo2021} and \cite{Pang2022}, respectively. 

While \nmma \ and \ztfrest \ enable powerful analysis, they still require manual input in order to initiate fitting new objects. Moreover, the existing framework does not have native support for conducting multiple analyses of the same object with different models. This online framework seeks to make use of the tools provided by \nmma \ and \ztfrest \ to automatically perform fits of new candidates to multiple models in a systematic fashion. This reduces the amount of time that needs to be spent initializing analysis of new candidates and allows for researchers to focus on choosing the most promising candidates for follow-up.

\subsection{Candidates}\label{sec:candidates}

\begin{figure}
\centering 
\includegraphics[width=3.3in]{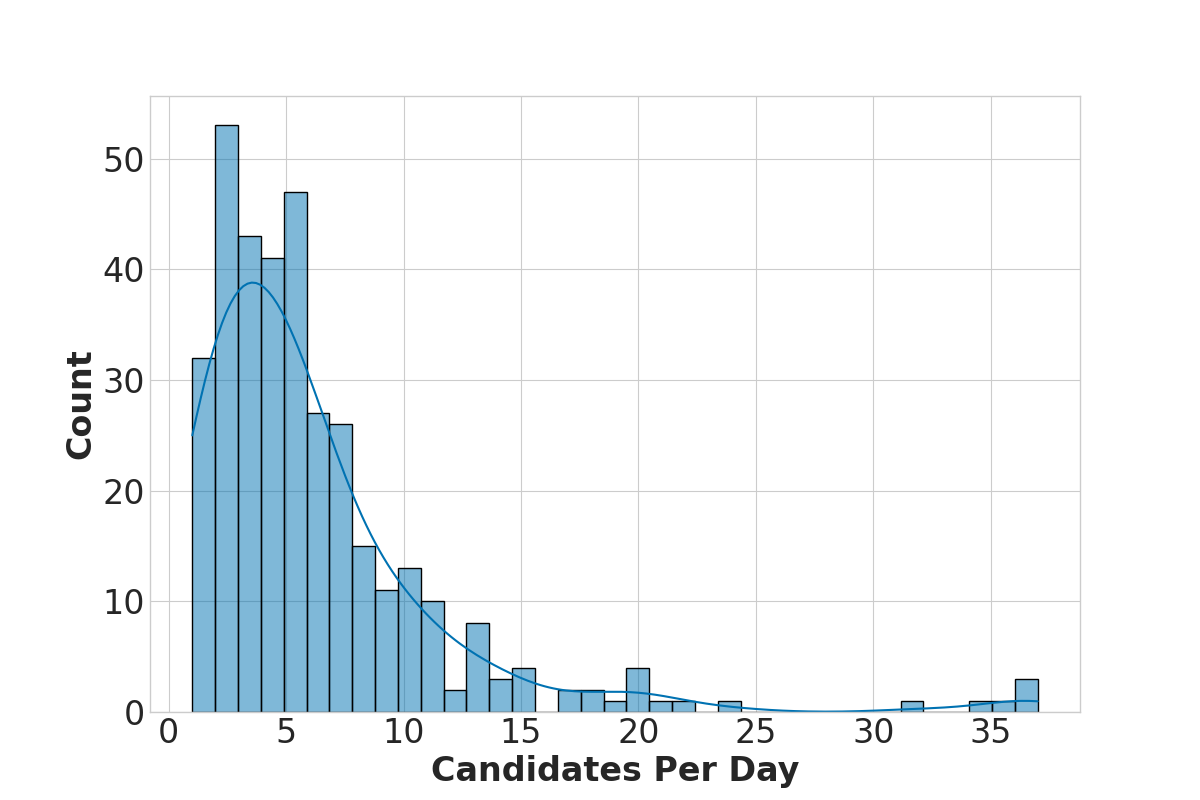}
\caption{A histogram of the number of candidates each day from July 2021 to October 2022. There were usually between 2 and 10 candidates analyzed each day. The plotted line represents a kernel density estimation of the histogram data. The median number of daily candidates was 5.}
\label{fig:numDailyCands}
\end{figure}

The goal is to identify rare transients like kilonovae and orphan afterglows (which occur when a GRB is not observed prior to the afterglow; this is discussed further in Section \ref{sec:models}) using multi-band, and potentially, multi-wavelength or multi-messenger data. The former is enabled by the use of alert streams of all sky optical surveys such as ZTF \citep{Bellm:19:ZTFScheduler,Graham2018,Masci2019,DeSm2018} and in the future, the Vera C.~Rubin Observatory's Legacy Survey of Space and Time (LSST; \citealt{Ivezic2019}). In the case of \ztfrest \ \citep{AnCo2021}, which relies on the ZTF alert stream \citep{Patterson2018}, the alert stream from both public and private ZTF surveys \citep{Bellm:19:ZTFScheduler} is used. Figure \ref{fig:numDailyCands} shows the distribution of daily candidates between July 2021 and October 2022 from this alert stream.

To identify rare and fast transients, we require a framework by which the nature of the transients could be determined based on time-series of photometric, multi-band lightcurves. In this way, their magnitude and color evolution, in particular those that are rapidly fading and/or reddening with no history of variability, can identify them as being of significant interest. For the most interesting objects, potentially spectroscopic classification and a well-sampled, multi-wavelength lightcurve to characterize the system is then desired. 

To do so, we perform parameter estimation of the lightcurves using \nmma, which is designed for Bayesian inference on multi-messenger signals \citep{DiCo2020,PaTe2021,TePa2021m,Huth2022,Pang2022} and has been used to analyze GW170817 \citep{DiCo2020}; AT2020scz, the shortest long GRB (LGRB) ever confirmed \citep{AhSi2021}; GRB 211211A \citep{Kunert2023}; and GRB 221009A \citep{2023arXiv230206225K}, a hyper-luminous GRB, amongst others.

We use the framework to obtain posteriors along with the log Bayesian evidence for the models presented below. A higher evidence value suggests that the model is a more plausible descriptor of the data. Conversely, a smaller value indicates that a model is less likely to accurately describe the data. Additionally, the log likelihood ratio of one model against another can be found by calculating the difference between the maximum log likelihood of the two models. A positive value of this odds ratio would suggest the first model is a better fit, whereas a negative value would imply the latter model fits the data better.

\subsection{Data Processing}\label{sec:MSI}
The bulk of our computing is done on the High Performance Computing (HPC) clusters maintained by the Minnesota Supercomputing Institute (MSI). 
Data from the Zwicky Transient Facility (ZTF) \citep{Bellm:19:ZTFScheduler,Graham2018,Masci2019,DeSm2018} alert stream is initially downloaded onto a system at the California Institute of Technology (CalTech) that is referred to as schoty.
The pipeline triggers a job every half hour on MSI to download data from the schoty directory that contains ZTF observations. Upon detecting new lightcurve files, the pipeline will trigger a series of fits to 4 different models, described in Section \ref{sec:models}, for each candidate.

After completing fits of all detected lightcurves, the results will be sent back to schoty. To make the fits easier to review, the pipeline will remotely execute a command to initiate a Slack bot script on schoty. This bot posts plots of each of the fits for the daily candidates as well as their posteriors and Bayes evidence values onto a channel on the GROWTH MMA Slack server. Since July 2021, this bot has successfully posted fits on a regular basis with only brief interruptions due to MSI or ZTF maintenance. 

Generally, this entire process takes on the order of 6 hours at the current level of detail. The rationale behind tuning the number of live points to reduce the total execution time is to allow for automating follow-up decisions in the future; this would require the fits to be completed some time before observing targets for the next night are set. There are several projects currently in-progress by members of the \nmma \ collaboration investigating methods by which the time required for fitting lightcurves can be reduced. One collaborator is investigating the use of machine learning on a collection of simulated lightcurves with known parameters to train an algorithm that would associate features of a lightcurve with certain parameter values. More generally, machine learning techniques represent an area of development that could enhance the speed and utility of the \nmma \ framework.

\begin{table*}
\centering
\begin{tabular}{|l|l|l|l|l|}
\hline
True Model              & Kilonova      & GRB Afterglow & Supernova & Shock Cooling \\ \hline
Kilonova                & 100\%         & 0\%           & 0\%       & 0\%            \\ \hline
GRB Afterglow           & 5\%           & 90\%          & 3\%       & 2\%             \\ \hline
Supernova               & 0\%           & 0\%           & 100\%     & 0\%              \\ \hline
Shock Cooling           & 0\%           & 28\%          & 36\%      & 36\%              \\ \hline
\end{tabular}
\caption{Results of model recovery and analysis with \nmma \ for the sample of simulated lightcurves described in Section \ref{sec:modelRecovery}. Each row is the distribution of what percent of each true model type was identified as the model listed in the column, such that each row will sum to $100\%$. The diagonal cells show what percent of each model was accurately identified as the correct model based on the Bayes evidence of the \nmma \ fits. Off-diagonal cells for each row will show what percentage of each model was incorrectly identified as a different transient type.}
\label{tab:model-recovery}
\end{table*}

\subsection{Lightcurve Models}\label{sec:models}

There are a number of astrophysical processes that produce fast transients. In this work, we will focus on a handful of extra-galactic astrophysical processes. For each model, both the explosion time and the distance is allowed to vary. We include Milky Way-like host extinction \citep{Fit1999} with $R_V = 3.1$ and $E(B-V)=0.1 \ mag$. 

The first are kilonovae, the optical counterparts to binary neutron star or neutron star-black hole system mergers generated by the $r$-process material that is produced in these events \citep{Me2017}. In our analysis, we use a \texttt{POSSIS}-based \citep{Bul2019} grid of kilonova models spanning the plausible binary neutron star parameter space \citep{DiCo2020}. There are four parameters, the dynamical ejecta mass $ M^{\rm dyn}_{\rm ej}$, the disk wind ejecta mass $M_{\rm ej}^{\mathrm{wind}}$, the half opening angle $\Phi$, and the observation angle $\Theta_{\rm{obs}}$. Internal code for the pipeline refers to this model as $\texttt{Bu2019lm}$.

The second process are GRB afterglows. We use \texttt{afterglowpy} \citep{RyEe2020}, an open-source computational tool modeling forward shock synchrotron emission from relativistic blast waves as a function of jet structure and viewing angle. The model parameters are the isotropic kinetic energy, $E_{\mathrm{K,iso}}$; the jet collimation angle, $\theta_c$; the viewing angle, $\theta_v$; the circumburst constant density, $n$; the spectral slope of the electron distribution, $p$; the fraction of energy imparted to both the electrons, $\epsilon_e$, and to the magnetic field, $\epsilon_B$, by the shock. This model is also called $\texttt{TrPi2018}$ in the code. 

The third process are supernovae. We use a model for SN Ib/c	supernovae \citep{LeNu2005} from the $\texttt{SNCosmo}$ package \citep{BaBa2016} with the stretch and scale set to match the intrinsic (dereddened, rest frame) $R$-band luminosity of SN\,1998bw at maximum light. The absolute magnitude is the free parameter with the largest effect on the quality of the fit. This is referred to as the $\texttt{nugent-hyper}$ model within the pipeline.

The fourth process are \textit{shock cooling} supernovae using a model that follows \cite{PiHa2021}. After shock breakout, the radiation of shock heated material expands and cools, known as shock cooling emission. The model considers extended material with mass $M_e$ and radius $R_e$, which is imparted with an energy $E_e$ as the shock passes through it. Internally, this is called the $\texttt{Piro2021}$ model. Generally, this paper refers to the $\texttt{nugent-hyper}$ model as SNe Ib/c or simply supernovae, while the $\texttt{Piro2021}$ model is referred to as shock cooling or shock cooling SNe.

In addition to the three non-kilonova models presented here, there are several other source categories that can imitate kilonova lightcurves, such as M-dwarf flares and cataclysmic variables; modeling these sources with \nmma \ and integrating them into this pipeline is an area for future work.

\section{Results}
\label{sec:results}

We evaluate the reliability of \nmma \ in recovering the different types of transients discussed in Section \ref{sec:models} through the use of simulated lightcurves. We also highlight the performance of the pipeline for two notable examples from the literature. Following this, the general daily performance of the automated pipeline infrastructure is summarized.

\subsection{Model Recovery}\label{sec:modelRecovery}
An important consideration for the utility of the framework is the predictive power of the fitting - that is to say, how often it can accurately identify a candidate of a given type as such. Ideally, a kilonova would always be best fit by a kilonova model as compared to other analysed models, with the same being true of other transient event types. 

To evaluate this, we used an existing \nmma \ tool that generates an injection file of parameter values that fall within the defined prior space of the provided model. We then use this injection file to \nmma \ to create a simulated lightcurve. By using a kilonova model to create an injection, we can simulate a kilonova lightcurve that could be plausibly observed. From there, we can apply the same multi-model analysis described above to evaluate how well \nmma \ is able to recover the correct model.

We created a sample of $100$ simulated lightcurves for each model using the same priors used for this pipeline. To minimize the dependence on observation cadences, we imposed a half day cadence in observations for all lightcurves and considered data from a single ZTF filter, the $g$-band. A realistic detection limit of $21.5$ magnitudes was also imposed; all detections above this limit up to $21$ days from initial detection was considered in the analysis. We then fit each lightcurve in the sample to each of the analysis models and evaluated the relative quality of the fits. This is accomplished by calculating the odds ratios of the true model for each lightcurve and the three other models. If all three of these odds ratios are positive, that indicates that the true model is the best fit for the lightcurve.

After noticing a high rate of failure for fitting shock cooling lightcurves, further investigation revealed that a majority of the shock cooling lightcurves only had a handful of detected data points, resulting in the analysis failing due to insufficient data. This prompted the addition of a method to enforce minimum observing conditions for injected lightcurves; a flag was added such that, if an injection file produced a lightcurve that did not have at least 3 detections within the first 3 days of observation, the injection was regenerated until one was produced that met the defined observing condition. In the case of the shock cooling model, this resulted in a noticeable difference in the distribution of the physical parameters associated with the model, implying that there are certain areas of the parameter space that correspond to shock cooling that would not be observable with a limiting magnitude of $21.5$.

We found that kilonova lightcurves were accurately identified $100\%$ of the time; this high rate of model recovery is likely due to the even and frequent cadence of the observations in the data. The characteristic rise of a kilonova is not necessarily likely to be observed in realistic observing conditions due to its short duration, but its presence in this sample is another potential factor causing the identification of kilonova to be so reliable. SNe Ib/c were also identified correctly $100\%$ of the time, which is most probably a confluence of the simplicity of the $\texttt{nugent-hyper}$ model and characteristic lightcurve shape. 

GRB afterglows are the first model with a sub-$100\%$ recovery rate, being correctly identified $90\%$ of the time. The $10\%$ of incorrect identifications are spread across the three other models, with $5\%$ being identified incorrectly as kilonovae and $3\%$ and $2\%$ being identified as SNe Ib/c and shock cooling SNe, respectively.

The shock cooling model had the lowest rate of recovery, with only $36\%$ of lightcurves being correctly identified as such. An equal number of shock cooling SNe were misidentified as SNe Ib/c, and the remaining $28\%$ were incorrectly identified as GRB afterglows. The summarized results of this model recovery can be found in Table \ref{tab:model-recovery}.

\subsection{Time Dependence of Model Recovery}\label{sec:modelRecoveryTime}
\begin{figure*}
    \centering
    \includegraphics[width=1\linewidth]{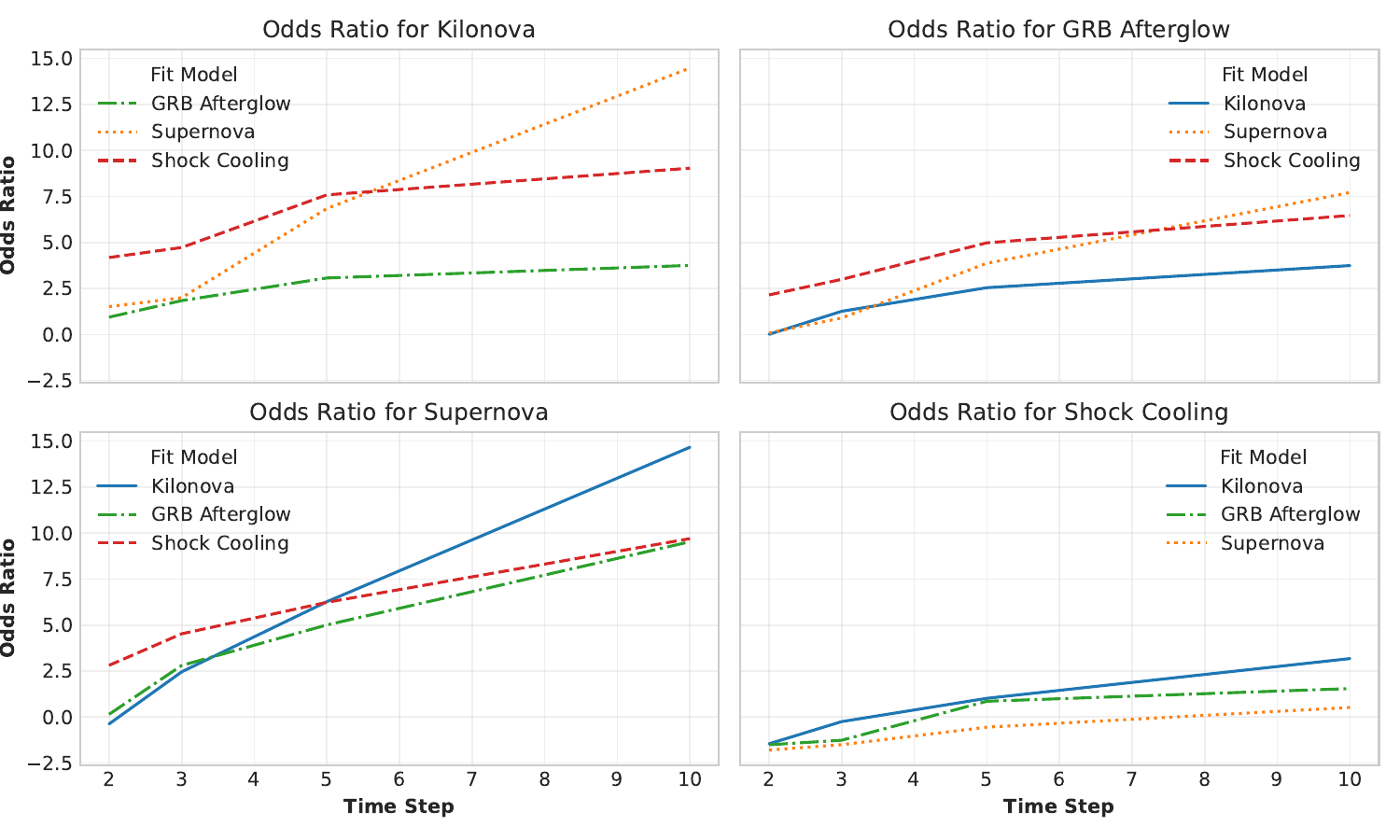}
    \caption{The median odds ratios of simulated lightcurves over time. Each subplot corresponds to the sample of lightcurves generated with the model listed in the subplot title (i.e., the true model). The different lines in each subplot represent the median odds ratio over time between the Bayes evidence of fits to the true model and fits to the other models. Generally, all models show a trend of odds ratios increasing over time, suggesting that \nmma \ analysis more strongly prefers the correct model as more data is considered.}
    \label{fig:odds_ratios}
\end{figure*}

A complimentary topic of interest is how the quality of fitting might change over time, especially at early times where there is a paucity of data. To investigate this further, a similar process was used as described in Section \ref{sec:modelRecovery}, but with the added complexity of performing a series of successive fits on each lightcurve with increasing amounts of data in consideration. In other words, The same series of fits would be conducted with the first 2 days of data from each lightcurve, then for the first 3 days of data, and so on going out to 10 days from the initial observation of each lightcurve. A minimum of 2 days was selected to allow for a sufficient number of detections such that \nmma \ could complete its analyses. This generates a large volume of analyses, over $10,000$ for a sample of $100$ lightcurves of each of the models. 
This, in turn, produces a large volume of data, both in total size and number of files, so the results are consolidated into a $.csv$ file with the best fit lightcurve and its accompanying parameters for each analysis. 

Upon initial evaluation of the rate of correct identifications over time, it became apparent that the kilonova model was being identified correctly essentially $100\%$  of the time, even when only provided the minimum number of data points. This could be due to a number of factors: as mentioned above, by observing kilonova early on, the lightcurve captures their characteristic rise, which may result in consistent identification early on. Most data from ZTF will be far less consistent in its observation cadence and it's unlikely that it will be possible to capture the early rise of kilonova with ZTF, so this set of lightcurves is not necessarily informative of how well \nmma \ will be able to perform under realistic scenarios. Another possibility is simply Occam's Razor - the kilonova model is less complex in comparison to the GRB model, so it is possible the kilonova is preferentially chosen by the Bayesian inference performed by \nmma ; we are actively investigating the potential source of this. 

Regardless, in order to evaluate the performance of \nmma \ in a more realistic manner, we make use of a feature already present in \nmma, which samples lightcurves on ZTF-like cadences. It also simulates variance in the ZTF limiting magnitude each night. To account for these changes, the observing condition for generation was loosened to only require a minimum of 3 observations within 5 days of the initial observation; this represents a source that is well-observed by ZTF but still plausible. Even with this relaxed requirement, some models saw a significant increase in the number of attempts required in order to generate a valid lightcurve. This was most apparent with the shock cooling model, which could require upwards of 200 attempts before generating a lightcurve that met the conditions. This reinforces the notion that areas of the parameter space, while physically possible, may result in objects that are unlikely to be observable by ZTF.

From this new sample of lightcurves, the Bayes evidence was aggregated at each time step for each combination of injected lightcurve and fit model. The median value of the Bayes evidence for each model at each time step was taken, and then the odds ratio was calculated between the true model and the other models in consideration. This can be seen in Figure \ref{fig:odds_ratios}. The primary takeaway from this metric is the increase in the median odds ratio over time for all models; with additional data, objects are better fit by their true model than by other models. This metric also suggested that all 4 models were fairly consistent in being accurately identified from early times, with the exception of the shock cooling model. However, evaluating the results by finding the model with the highest odds ratio for each lightcurve at each time step demonstrated a different result.

The kilonovae were still identified fairly reliably with the minimum amount of data (2 days), suggesting that the characteristic rise may play a role in the rate of identification. There is a small, temporary dip in the rate of correct identifications when considering a maximum time of 3 days. This could be a result of small-number statistics, as this analysis only contains 100 lightcurves for each model, but it could also indicate there's an intermediate amount of data where kilonovae can appear to resemble supernovae analytically.

Supernovae demonstrated a high rate of misidentification with the minimal amount of data but rapidly increase in identification with an additional day of data, reaching $100\%$ identification by the time there was 5 days of data in consideration. It took somewhat longer for GRB afterglows to reach a majority being correctly identified, though it still demonstrates a positive trend. Regardless of the amount of data, all shock cooling SNe were misidentified, with a majority being classified as the standard SN model. 

The low rate of recovery for the shock cooling model suggests that, while they were incorrectly identified, the individual odds ratios between the best fit and the other fits are fairly small. The median odds ratio values in the shock cooling subplot in Figure \ref{fig:odds_ratios} supports this, with none of the odds ratios exceeding 2.5 prior to the tenth day of observations, though they do trend positive. This suggests that there is some lower cutoff for the odds ratio below which one cannot confidently assert the likelihood of one model over another.

While the rate of identifying kilonovae was fairly high, even with the minimal amount of data, it's worth noting the number of other candidates that were incorrectly identified as kilonovae. With the minimum amount of data, there were a larger proportion of lightcurves that were incorrectly identified as kilonovae compared to the number of correctly identified kilonovae. It's worth noting that this is a sample of an equal number of each model; the actual number of kilonovae in comparison to other transient types would be significantly lower with a more realistic sample of transients. 

Inspecting the odds ratios between the best and next best fits for both the true kilonovae and the imposter kilonovae, many of the imposter kilonovae had odds ratios that trended worse than the odds ratios of the true kilonovae, which represents a possible avenue for evaluating the merits of sources in the event multiple objects in a given night are best fit by a kilonova model.

\begin{figure*}
    \centering
    \includegraphics[width=1\linewidth]{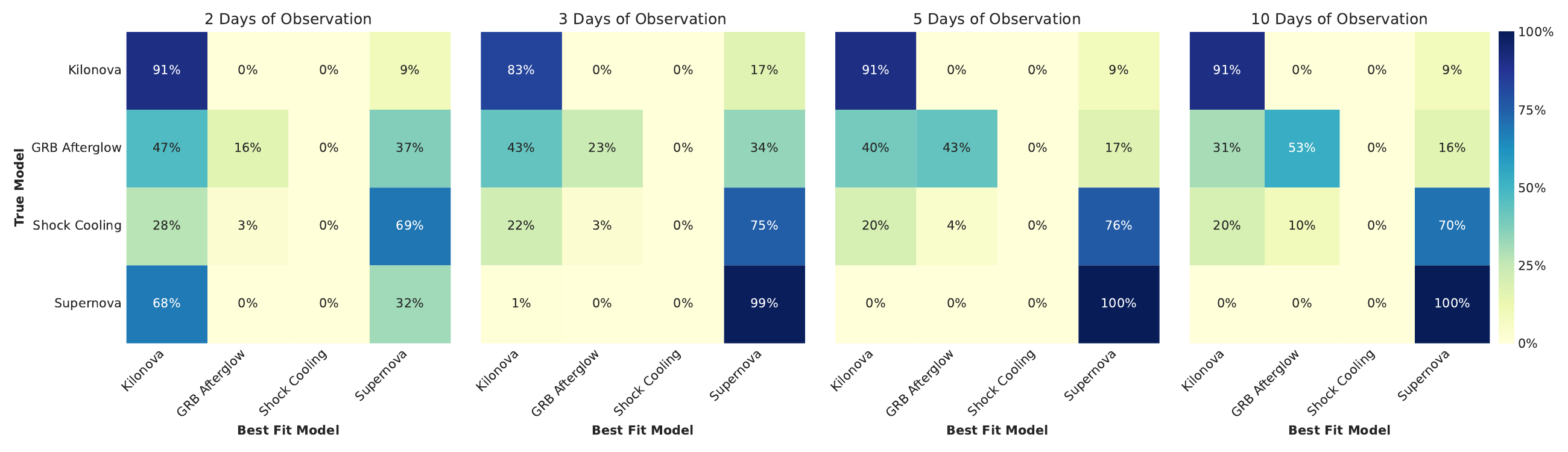}
    \caption{A collection of confusion matrices denoting the distribution of the best fit models versus the true models of simulated lightcurves described in Section \ref{sec:modelRecoveryTime}. In each confusion matrix, the diagonal elements represent correct identifications and the off-diagonal elements are incorrect identifications; each row should sum to $100\%$, representing the collection of simulated lightcurves for that model, and each column denotes the relative percent of lightcurves identified as being best fit by a given model. Each matrix corresponds to fits done with data for each lightcurve from their initial detection up to the number of days listed in their respective titles.}
    \label{fig:confusion_matrix}
\end{figure*}

\subsection{Multi-Armed Bandit}\label{sec:mab}
Given the time-sensitive nature of follow-up observation, we explored using concepts from statistics to optimize decision making. We have implemented a multi-armed bandit (MAB) algorithm, which is a strategy that uses a defined reward to make the optimal selection from a number of different possible choices. \cite{slivkins2024introduction} provides an extensive introduction to various MAB algorithms.

In our case, the MAB selects the most likely kilonova from an ensemble of candidates. For a given observing scenario, we include nine candidates: one kilonova, one GRB afterglow, and one supernova, generated using the same models as detailed in previous sections. 

Lightcurves for each observing scenario are sampled on a ZTF-realistic cadence using the same functionality described above, though we impose a broader requirement of 8 $g$-band detections within 18 days of the first detection to allow for the MAB to be run over a longer period of time. Again, we use \nmma \ to fit each lightcurve to each model; the results of these analyses are then used to evaluate a reward function.

The reward is defined as the odds ratio of the kilonova model and the best non-kilonova model as well as an upper confidence bound (UCB) term that allows for exploration of the parameter space. Essentially, this term prevents the algorithm from exclusively selecting a specific object unless the odds ratio continues to improve with additional observations or is markedly higher than odds ratios for other objects at early times. The reward equation is given by

\begin{equation}\label{eq:reward}
    R_n(t) = \left( \text{L}(t)_{n,KN} - \text{L}(t)_{n,non-KN}\right) + \sqrt{\frac{2 \text{ln}(t)}{N_n(t)}}
\end{equation}

where $t$ is the timestep being evaluated and the subscript $n$ differentiates between candidates. The leftmost term is the odds ratio of the kilonova model fit and the next best model fit. The rightmost term is the UCB, where the numerator is two times the natural log of the current timestep and the denominator is the number of times that object $n$ has been selected by the MAB.

\begin{figure}
    \centering
    \includegraphics[width=3.3in]{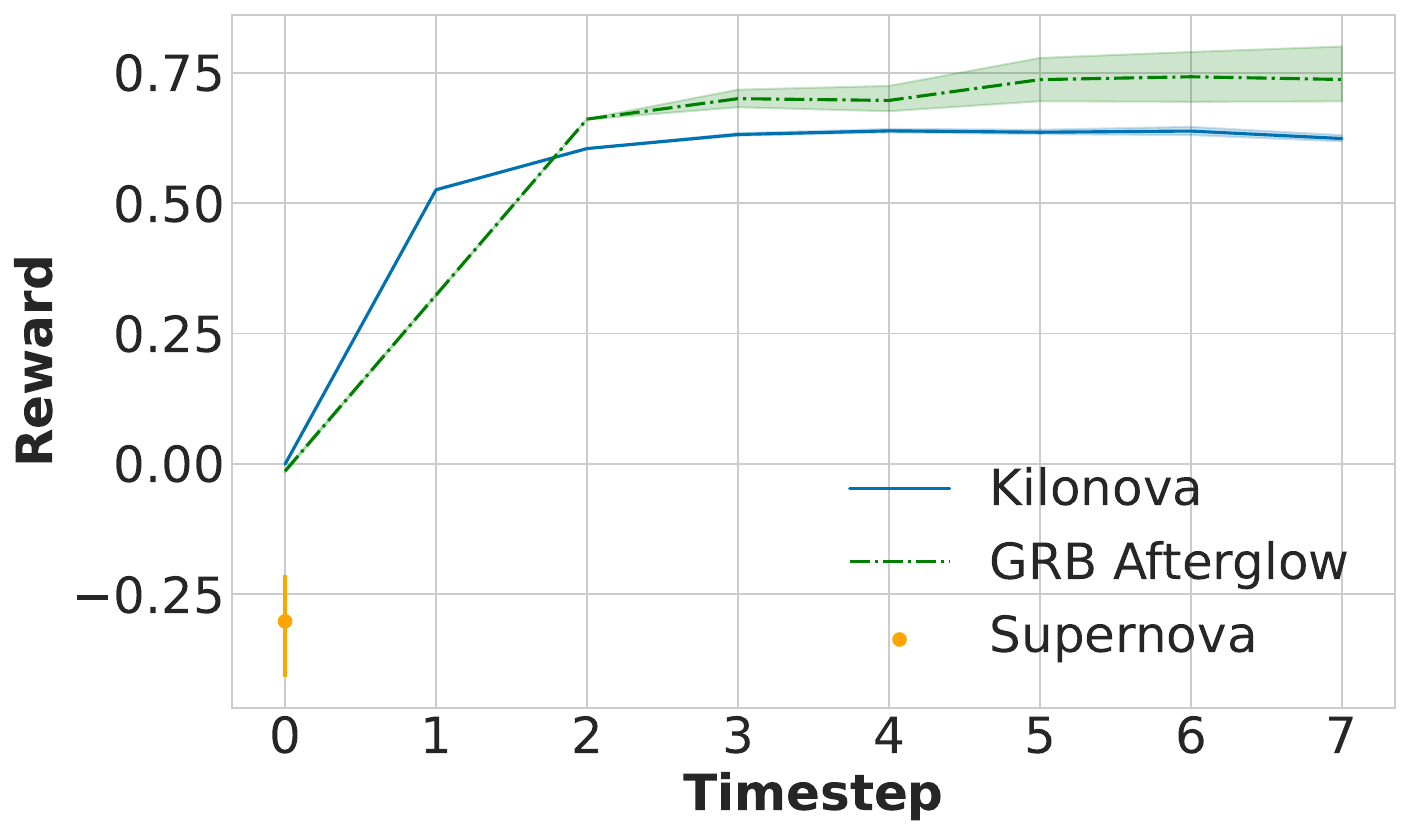}
    \caption{The MAB reward, as calculated by Equation \ref{eq:reward}, for the ensemble of observing scenarios, where the lines show the average value of the rewards for objects observed at that time step and the shaded areas represent the 95\% confidence interval. No supernovae were selected for further observation by the algorithm, so the rewards for supernovae are plotted as a single point at timestep 0.}
    \label{fig:MABReward}
\end{figure}

Generally, the reward will be positive if the kilonova model is the best fit and it will be negative if it is not. If multiple candidates have a positive reward, the candidate with the highest reward will be selected. In this case, ``selection'' refers to being chosen for follow-up observation. The candidate with the highest reward is chosen for observation during the next observing period. Ideally, the MAB will quickly identify the correct candidate and continue to observe the kilonova, while the UCB term will allow for the consideration of other candidates if the odds ratios between the best candidates are similar.

Similar to Sec.~\ref{sec:modelRecoveryTime}, each lightcurve is simulated in its entirety before being concatenated into an initial set of observations, namely when there are a minimum of three observations for each candidate, representing timestep 0. After analyzing these initial lightcurves, the reward is evaluated for each candidate. For the selected candidate, any observations made in the interval spanning the next day from the full lightcurve are added to the concatenated lightcurve and the analysis is conducted with the additional data. 

To evaluate the performance of the MAB algorithm, we ran the algorithm on 100 observing scenarios, using a timestep of 2 days for calculation of the reward function and object selection and a total of 7 timesteps. We found that the MAB was fairly effective in selecting kilonova candidates early and often. The kilonova was correctly selected in all 100 observing scenarios at the first time step; at the final time step, the kilonova was correctly selected in 93 of the 100 observing scenarios. Across all timesteps, kilonovae were selected for observation $92.6\%$ of the time. In 80 of the 100 observing scenarios, the kilonova was selected at every timestep.

At all timesteps where the kilonova was not selected, the GRB afterglow was selected instead; there were no cases of supernovae being selected at any timestep. The rewards for the objects observed at each timestep are shown in Fig.~\ref{fig:MABReward}. This figure contextualizes the lack of additional supernova observations given the significantly lower average reward when compared to the other candidate types.

While the average GRB afterglow reward exceeds the average kilonova reward at timestep 2 and beyond, the rate of GRB afterglow selection peaked at $12\%$ (also at timestep 2), with an average selection rate of $7.4\%$ across all timesteps. Because of the relatively low selection rate for GRB afterglows, those that are selected would tend to have higher reward values than the average kilonova reward. 

Even in cases where a GRB afterglow was chosen, the kilonova was still selected for a majority of the timesteps. Of the 20 observing scenarios where the kilonova was not chosen every timestep, the GRB afterglow was selected more than the kilonova only 7 times, and there were no scenarios where the GRB afterglow was selected at every timestep. For GRB afterglows that were selected at least once, the average number of selections (out of a possible 7) was 2.6, as compared to an average of 6.5 selections for all kilonova selected at least once. This suggests that, even for scenarios where differentiating between a kilonova and an imposter via photometric analysis might be difficult, the MAB algorithm is unlikely to select the incorrect candidate for follow-up observation more than the true kilonova. 

This initial study of the utility of MAB algorithms for optimizing follow-up observation is promising, and we plan to integrate MAB functionality into the \nmma \ framework to allow for future analysis. Particularly, we plan to evaluate its performance across larger samples of candidates, with additional classes of eFOTs, and using data from multiple filters.

\subsection{ZTF20abwysqy}\label{sec:ZTF20abwysqy}
ZTF20abwysqy, also known as AT2020scz, is an afterglow that was identified as the optical companion of the GRB 200826A, which is noted in \cite{Tomas2021} as the shortest LGRB discovered, suggesting it may straddle the minimum conditions for which a successful collapsar can occur. While this was originally detected by Fermi, follow-up observations with ZTF allowed for the identification of the optical counterpart \citep{AnCo2021}. 

As detailed in \cite{AnCo2021}, \ztfrest \ can create lightcurves both from ZTF alerts and forced photometry. These lightcurves can then be fit by the pipeline. While X-ray observations with Swift would have ultimately identified ZTF20abwysqy as an afterglow, the use of forced photometry with \ztfrest \ allowed it to be recognized as a rapidly-fading source more promptly and motivated follow-up observation.

\begin{figure}
    \centering
    \includegraphics[width=3.3in]{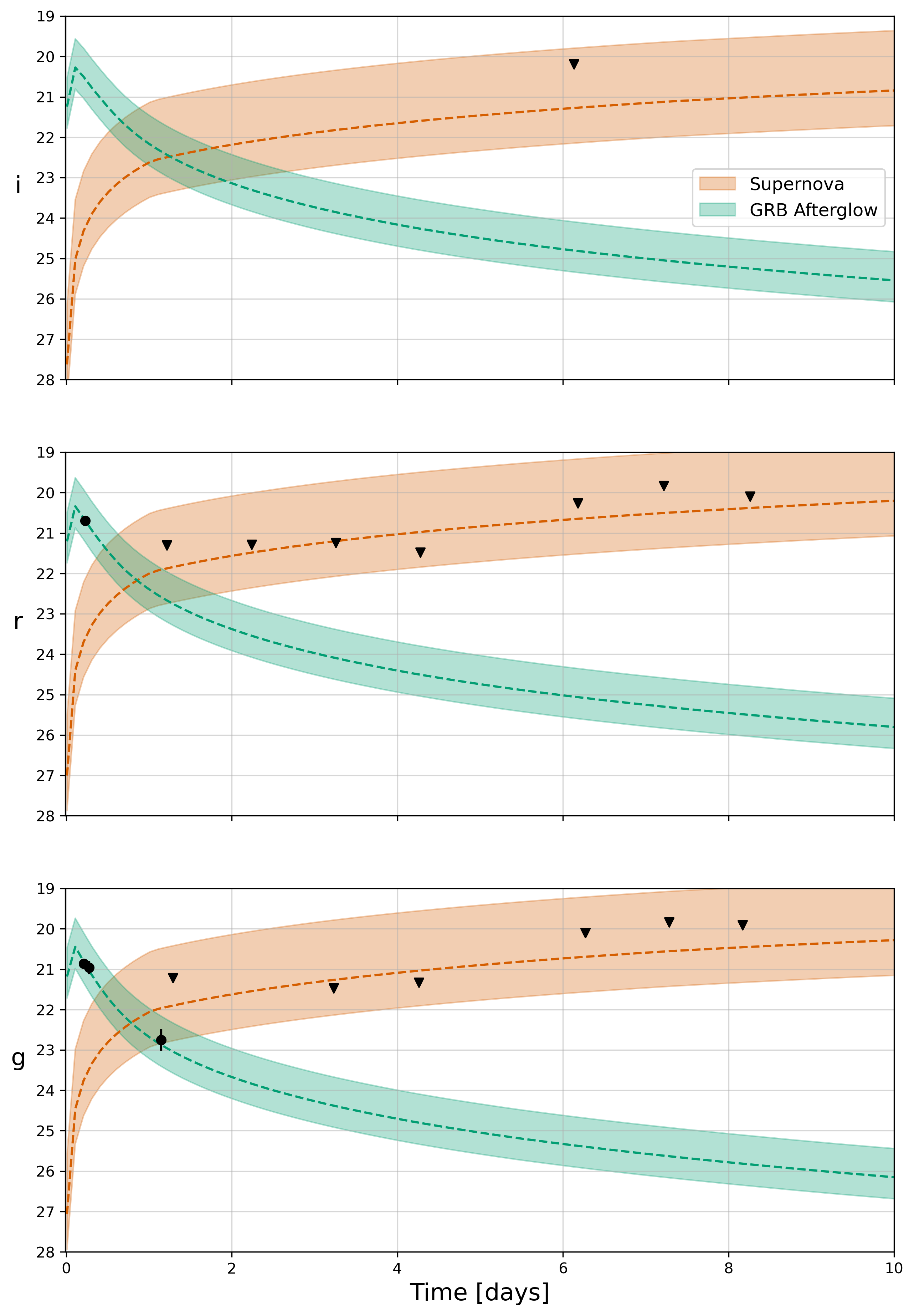}
    \caption{The lightcurve of the ZTF20abwysqy and analyses of the object fit to the supernova and GRB afterglow models; non-detections are denoted as triangles, while detections are represented by circles. The colored lines for each model are the best fit, and shaded regions correspond to one standard deviation in estimated luminosity distance among the samples calculated for a given model fit. This object was initially suspected to be a supernova, but additional observations suggested it was the optical companion of a GRB. Its classification as a such was later confirmed in a follow-up study.}
    \label{fig:ZTF20abwysqyModels}
\end{figure}

\begin{figure}
    \centering
    \includegraphics[width=3.3in]{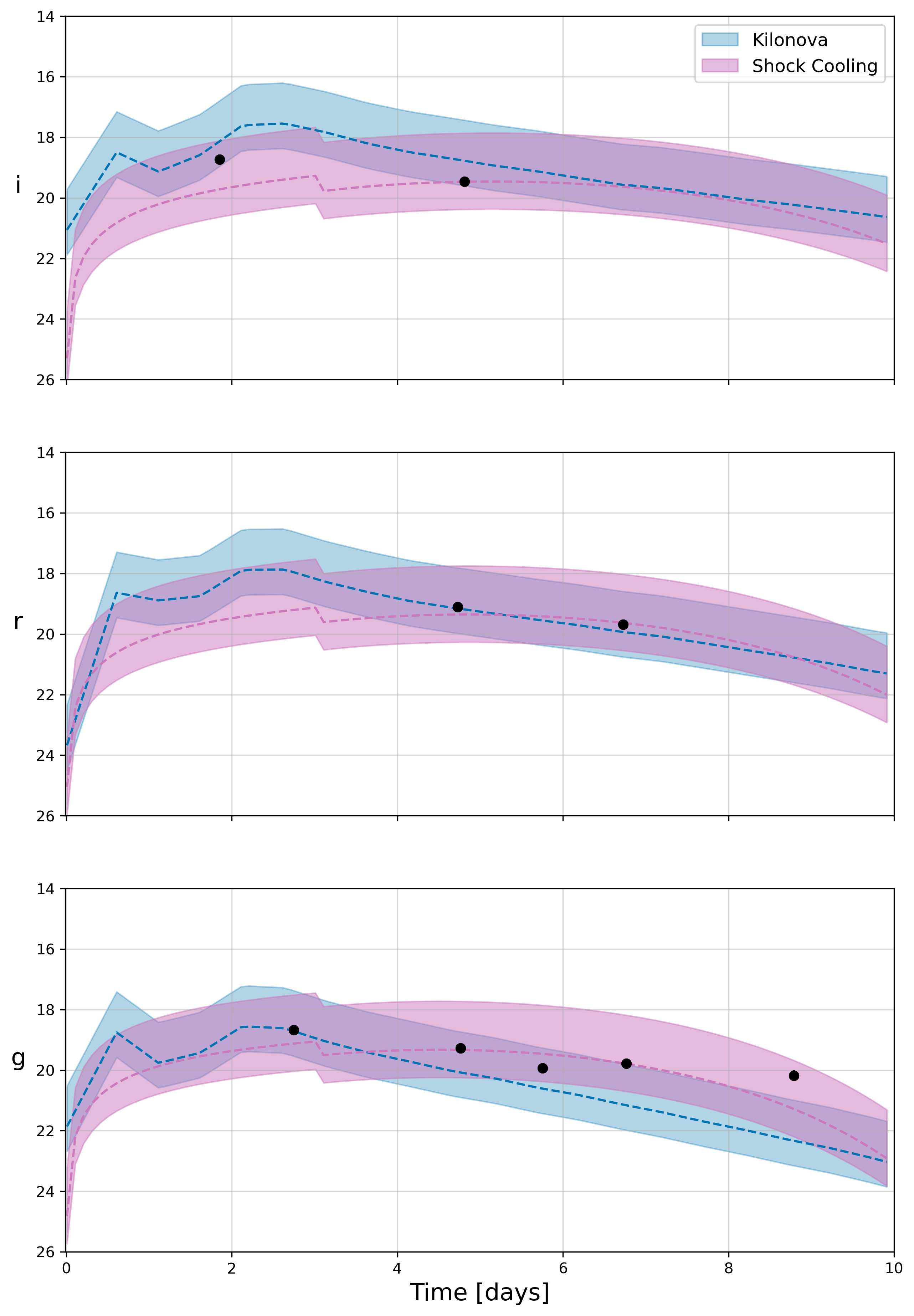}
    \caption{The lightcurve of ZTF21abotose and fits to the kilonova and shock cooling supernova models. As in Figure \ref{fig:ZTF20abwysqyModels}, the lines for each model show the best fit, and the shaded regions represent one standard deviation of the estimated luminosity distance. This object was difficult to classify based on photometric analysis alone, but spectroscopic analysis ultimately confirmed its classification as a Type IIb supernova.}
    \label{fig:ZTF21abotoseModels}
\end{figure}

Analysis of ZTF20abwysqy was conducted using this pipeline for both the supernova and GRB afterglow models; the results of these analyses is shown in Figure \ref{fig:ZTF20abwysqyModels}. Visual inspection of the fits to ZTF20abwysqy suggests that the non-detections cannot entirely rule out the possibility of it being a supernova, there is a notable difference in the Bayes evidence values between the two. The odds ratio of the Bayes evidence values between the GRB afterglow model and the supernova model is $19.77$, which seems to favor the GRB afterglow model in comparison to the supernova model. In comparison, the simulated lightcurves for GRB afterglows, shown in the upper-right subplot in Figure \ref{fig:odds_ratios}, had a median odds ratio of around $7.5$ with roughly the same amount of observations available for analysis (though the fits of ZTF20abwysqy benefit from limiting magnitude measurements in the $i$ and $r$-bands as well as an additional data point in the $r$-band). This suggests that this pipeline and \nmma \ would have supported the idea of ZTF20abwysqy being a GRB afterglow rather than a supernova at the time these observations were occurring, which could have motivated additional follow-up observation.

\begin{figure*}
\includegraphics[width=7in]{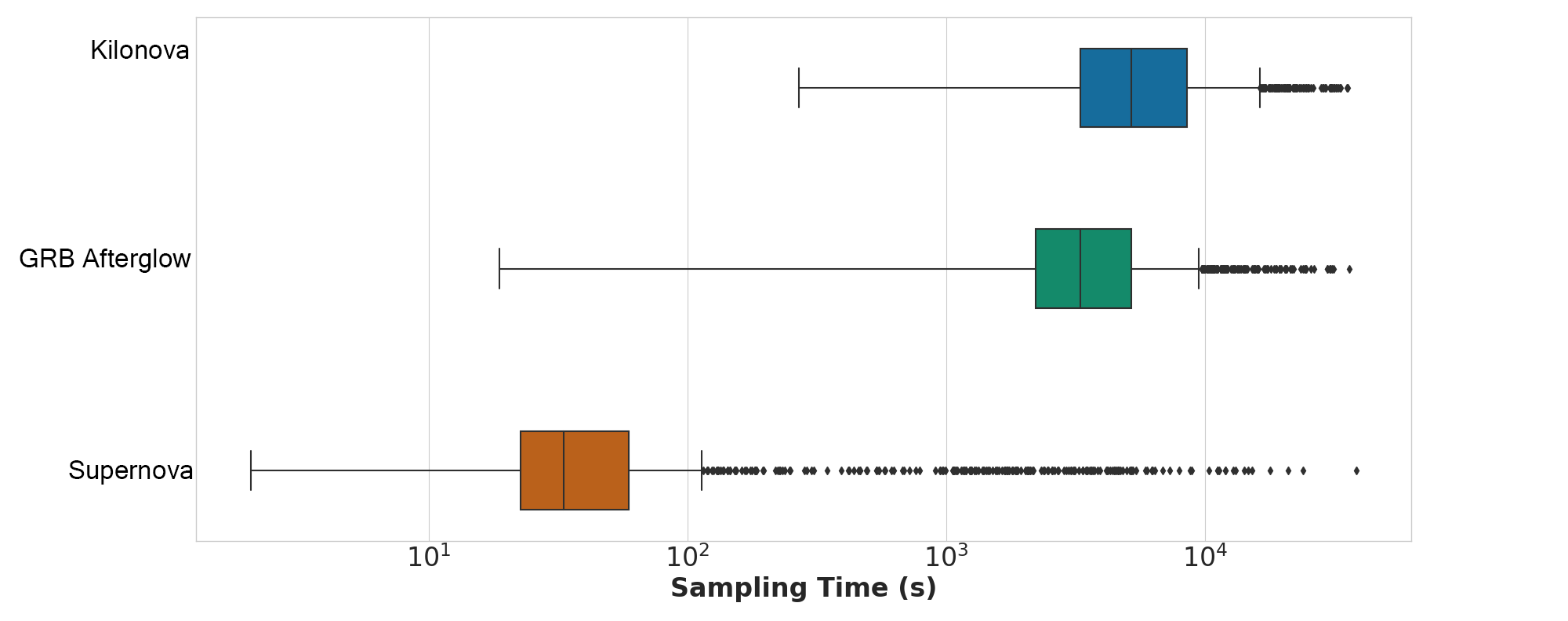}
\centering
\caption{Distribution of fit sampling times by model. The x-axis corresponds to the time in seconds required to complete an analysis of a given model. Each box represents the inner quartile range of the data, with the line inside this box being the calculated median, and outliers are plotted as individual points. Note the log scaling of the x-axis; the median sampling time for the supernova model is roughly two orders of magnitude less than the median values of the GRB and kilonova models, due in part to the relative simplicity of the supernova model.}
\label{fig:samplingTimeDist}
\centering
\end{figure*}

\subsection{ZTF21abotose}\label{sec:ZTF21abotose}
The primary goal of \ztfrest \ is to find kilonovae and GRBs, but there are cases where other transients might appear to match the form of a kilonova or GRB. Objects like ZTF21abotose exemplify both an "impostor" among candidates and a potential avenue for expansion of the pipeline. 

From initial ZTF observations, ZTF21abotose, also referred to as SN2021ugl, originally appeared to be a kilonova, but spectroscopic analysis of the object revealed it was actually a Type IIb supernova \citep{2021TNSAN.220....1R}, which occurs as a result of core collapse of a massive star.  More specifically, it was the signature of cooling following a shock breakout. When the core of a star collapses in on itself, a shock propagates outward from the inner core. Once this shock makes it to the surface of the star, it creates a spike in luminosity separate from the ongoing supernova \citep{2007PhR...442...38J, Campana2006, 1995ApJ...450..830B}. 

In the past, it has been challenging to observe shock breakout events, as they occur at the earliest stages of a core collapse supernova. Observations of shock breakouts have largely been a result of serendipitous observation at the time of explosion \citep{2008Natur.453..469S}. Now, with automated all-sky surveys like those conducted with ZTF, the observation of shock breakouts have become much more prevalent. In fact, they represent one of the primary classes of transients that are detected by \ztfrest \ outside of the intended targets. 

Analysis of ZTF21abotose with this pipeline was conducted for both the shock cooling and kilonova models, the results of which are shown in Figure \ref{fig:ZTF21abotoseModels}. Generally, there is significant overlap between the best fits for both the kilonova and shock breakout models; only the final observed data point in the $g$-band falls outside the projection of the kilonova model fit, though this does not necessarily rule it out as a potential kilonova. Inspection of the associated odds ratio between the Bayes evidence values of the shock breakout and kilonova models is $1.55$, which is notably lower than that of the one found for ZTF20abwysqy above. While the odds ratio for ZTF21abotose slightly favors the shock cooling model over the kilonova model, this analysis does not strongly support one model over the other.

Though detections of Type IIb SNe and other transients that resemble kilonovae and GRBs are troublesome for our current objectives, their prevalence in the ZTF data suggests a potential for \ztfrest \ to be employed in aiding in the search of other transient objects.

\subsection{General Performance} 
\label{performance}
Between July 2021 and October 2022, the pipeline executed on 354 days with a total of 2,149 daily candidates, of which 517 (roughly one quarter) were unique candidates. Candidates were observed for an average of just over 4 nights, with the maximum number of observations being 16. The average (median) number of candidates analyzed each day was 6.1 (5.0), and the average (median) number of unique candidates was 2.8 (2.0). A histogram of the number of daily candidates during the analysis period is shown in Figure \ref{fig:numDailyCands}.

Of the four models implemented on the MSI system, the shock cooling model remained non-functional for the duration of the analysis period. The precise reason for this is unclear since it is possible to run fits with the shock cooling model locally and investigation is ongoing. Overall, 71\% of candidates had a successful fit to at least one model. That remaining 29\% could have been not fit for several reasons. Some candidates do not meet the filtering requirements we impose, and even if they pass the filtering requirements, it is possible that \nmma \ may not be able to converge on a solution. 

During the course of pipeline development, a manual time limit of 8 hours was imposed on all model fits so as to allow fits to be reviewed before the next night of observing. Figure \ref{fig:samplingTimeDist} shows the distribution of sampling times on a per-model basis, where sampling time refers to the amount of time in seconds required to complete an analysis using \nmma; there is a noticeable difference in the time required to fit the supernova model as compared to the other models. The time required to complete analysis of a given candidate is a meaningful metric by which to evaluate the feasibility of using this framework to inform follow-up observation. Excessively long sampling times would result in analyses not being completed prior to the next night of observing, reducing the utility of the framework. However, the results of the analysis period suggests the framework is able to complete most individual analyses within an hour.

\begin{figure}
\centering 
\includegraphics[width=3.3in]{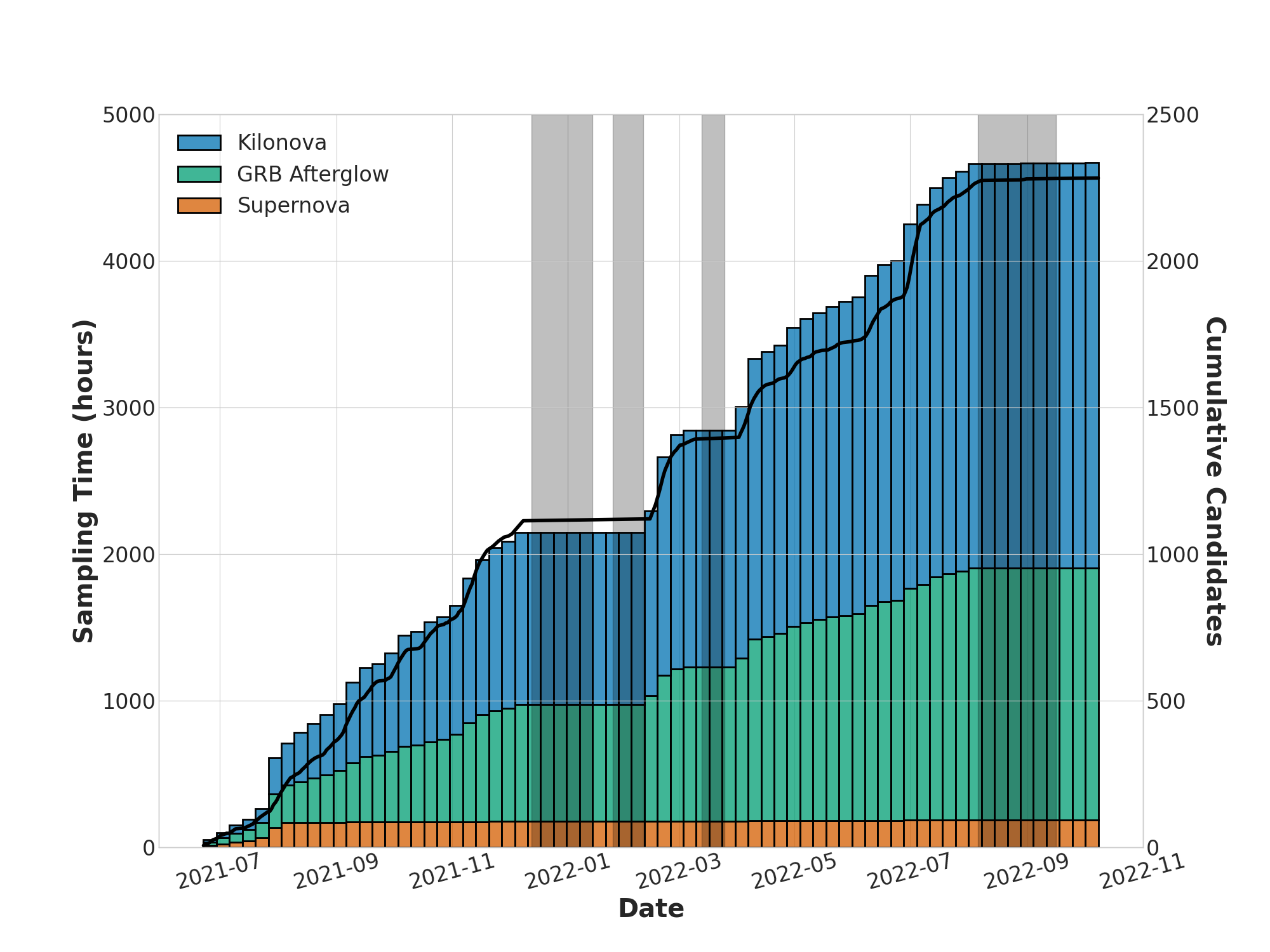} 
\caption{The bar plot (left axis) represents the cumulative sampling time, in hours, on MSI between July 2021 and October 2022. The total sampling time was roughly 4670 hours. The line (right axis) represents the cumulative number of candidates, including candidates that were analyzed on more than one day. The greyed-out areas on the plot represent periods where the pipeline was not consistently receiving new data. For the leftmost 3 instances, this was due to ZTF being down for maintenance or service. The rightmost instance was a result of issues with the connection between MSI and schoty.}
\label{fig:cumFitTime}
\end{figure}

In total, the pipeline used roughly 4,760 computing hours during the analysis period, which corresponds to an average (parallel) computing time of just over 13 hours per day. The supernova, GRB afterglow, and kilonova  models made up 4\%, 37\%, and 59\% of the computing time, respectively. Figure \ref{fig:samplingTimeDist} shows that the median computing time for the supernova model fits is on the order of minutes, whereas the other two models generally take almost two orders of magnitude longer to complete. The supernova model is a less complex model with fewer parameters, so \nmma \ requires much less time for likelihood evaluations and converges faster.

As shown in Figure \ref{fig:cumFitTime}, the cumulative computing time tracked fairly closely with the cumulative number of candidates analyzed with the exception of the supernova model. Beyond the first month, the supernova model computing time grew very little, making up an increasingly small proportion of the cumulative computing time. This figure also demonstrates the framework's relatively low usage of computing, requiring an average of just under 13.5 hours of computing time per day, which is generally run in parallel so as to complete daily analyses within a few hours.

\begin{figure}
\centering 
\includegraphics[width=3.3in]{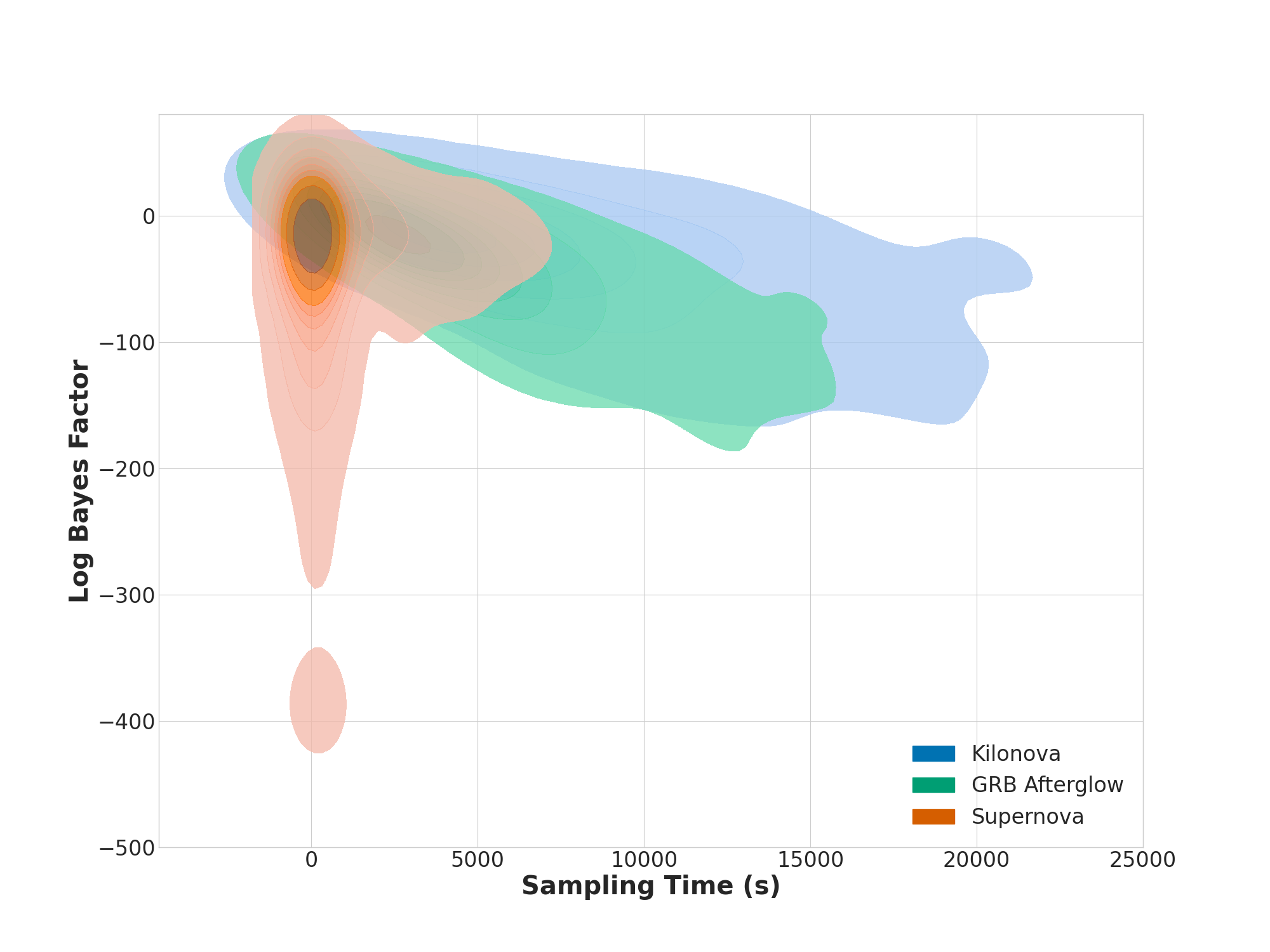}
\caption{A contour density plot of the sampling times compared to the Bayes evidence by model. Generally, the supernova model completed fitting within minutes, whereas the median sampling times for the kilonova and GRB afterglow models were on the order of hours. However, the supernova model demonstrated a much larger range in values for the Bayes evidence compared to the other models. Note that, while the contours may visually extend past the x-intercept, no fits took negative time to complete. This is also true for the y-intercept - the upper limit of the Bayes evidence is 0.}
\label{fig:samplingTimeBayes}
\end{figure}

Each of the models had similar maximum and minimum Bayes evidence values. However, the distribution of these Bayes evidence values differ. Figure \ref{fig:samplingTimeBayes} demonstrates the relationship between sampling times and the resulting Bayes evidence. There was not a particularly strong correlation between the Bayes evidence of supernova model fits and their sampling times, but significantly more of fits of this model had very negative Bayes evidence values. This makes sense; while the lower dimensionality of the supernova model resulted in shorter sampling times, the other two models are more likely to produce better Bayes evidence values because there are more parameters by which to fit the data; additionally, the \ztfrest \ algorithm is designed for fast transients, which supernova are not.

The kilonova and GRB afterglow models had a similar distribution of sampling times and Bayes evidence values; both indicate a general trend of longer sampling times resulting in more negative Bayes evidence values, with the distribution of sampling times for the kilonova model extending slightly further than that of the GRB afterglow model.

\section{Discussion}
\label{sec:discussion}    

In this paper, we have described a useful extension of the \ztfrest \ pipeline that can be generalized for use with other data sets and potential future functionality. We have evaluated the general performance of this pipeline as well as the performance of \nmma. As O4 continues and we look to future observing runs, this will provide an opportunity to identify additional needs for enabling the automatic fitting of fast transient lightcurves to inform follow-up observations of intriguing sources.

The pipeline has demonstrated the ability to perform fits on a daily basis without additional oversight. That being said, there are a number of open questions and potential functionalities that can be developed for the pipeline.

One of the principal questions that must be answered as development continues is how comparison between different fits can be quantified in a meaningful way. In order to properly inform automated follow-up observations, we must investigate what constitutes a significant enough difference in quality of fits between models to motivate additional observation of a target. By expanding on the model recovery analyses discussed in Sections \ref{sec:modelRecovery} and \ref{sec:modelRecoveryTime} to generate a more expansive set of lightcurves, it should be possible to quantify this difference. 

The use of \nmma \ in this pipeline demonstrates its ability to perform comparative analysis of ZTF lightcurves to tentatively identify their transient type to motivate follow-up observation. Features of this pipeline would be beneficial to integrate into \nmma \ to increase its utility as a multi-purpose analytical tool for MMA science. \nmma \ currently supports combined analysis of multiple transient models on a single lightcurve, but work is being done to enable users to provide multiple models as part of a single command that will then perform independent analysis of the lightcurve to the models. By integrating features from this pipeline, \nmma \ could then rank the relative quality of the models in fitting the lightcurve. Extending this further, it would then be possible to develop a built-in method that performs this comparative analysis on multiple lightcurves and evaluate which is most likely to be the model of interest.

The work outlined in Section \ref{sec:modelRecoveryTime} also lays the groundwork for another addition to the \nmma \ framework. By generating a large sample of lightcurves for various common transient sources and using \nmma \ to analyze their relative fit quality as additional observations are considered, this could form the basis for a machine learning (ML) training set using one of the several ML libraries available for \texttt{Python} to optimize the selection of a candidate from a set of multiple lightcurves that would maximize the odds ratio for the transient type of interest to the user. This is a valuable prospect when one is limited in the number of candidates that can be triggered in a given night for follow-up observation.

During the course of this project, it became apparent that this body of analyses would be beneficial to make available to the broader collaboration outside of members of the Growth MMA Slack server. As a result, an extension to the pipeline was made such that the analyses are now made available on Fritz, which is an open-source platform for time domain astronomers. Along these same lines, work is being done to develop an \nmma \ API to enable \nmma \ analyses to be requested an executed through Fritz. After adding the aforementioned features to \nmma \ motivated by this pipeline, they could also be made available through the API to enable users to leverage the power of \nmma \ in analyzing lightcurves and informing follow-up observations while using Fritz. Extending the API even further, it should be possible to automatically perform these comparative analyses on new objects as they're posted to Fritz. Adding these features to \nmma \ will be a boon to researchers and enable more productive time-domain astronomy.

Most of the outstanding questions are statistical in nature. With additional time and data, more definitive evaluations of fit quality will be able to be made automatically as we determine statistically significant measures. One consideration is the current level of manual review conducted prior to processing by the pipeline. As of this writing, only a handful of the most promising targets are considered. It may prove beneficial to relax this filtering process and allow the pipeline to evaluate less promising targets. Statistical analysis of the difference in fit quality between these two groups could reveal a measure that could be added to the pipeline to automate this filtering in the future.

\section*{Acknowledgements}

M.W.C acknowledges support from the National Science Foundation with grant numbers PHY-2308862 and PHY-2117997. Tyler Barna is supported through the University of Minnesota Data Science in Multi-Messenger Astrophysics (DSMMA) program by the National Science Foundation with grant number NRT-1922512.

Based on observations obtained with the Samuel Oschin Telescope 48-inch and the 60-inch Telescope at the Palomar Observatory as part of the Zwicky Transient Facility project. ZTF is supported by the National Science Foundation under Grants No. AST-1440341 and AST-2034437 and a collaboration including current partners Caltech, IPAC, the Weizmann Institute of Science, the Oskar Klein Center at Stockholm University, the University of Maryland, Deutsches Elektronen-Synchrotron and Humboldt University, the TANGO Consortium of Taiwan, the University of Wisconsin at Milwaukee, Trinity College Dublin, Lawrence Livermore National Laboratories, IN2P3, University of Warwick, Ruhr University Bochum, Northwestern University and former partners the University of Washington, Los Alamos National Laboratories, and Lawrence Berkeley National Laboratories. Operations are conducted by COO, IPAC, and UW.

The ZTF forced-photometry service was funded under the Heising-Simons Foundation grant \#12540303 (PI: Graham).

\section*{Data Availability}

Data regarding job completion statistics is available upon request; observational data reported in this article is publicly available on the Transient Name Server.


\bibliographystyle{mnras}
\bibliography{references}

\bsp
\label{lastpage}
\end{document}